
\magnification=1200
\vsize=22.5 truecm
\hsize=16 truecm
\voffset=0.5 truecm
\hoffset=0 truecm
\baselineskip 20 truept
\overfullrule=0 pt

\centerline{\bf Quantal Bifurcations in Rotational Spectra}
\centerline{\bf  of Some Odd Atomic Nuclei$^{(\star)}$}

\vfootnote*{
 This work  has been carried out within the scientific co--operation
agreement between ENEA and the University of Padova and partially supported by
the Ministero dell'Universit\`a e
della Ricerca Scientifica e Tecnologica (MURST).
\smallskip
\noindent
All correspondence and reprint requests should be addressed to:
Prof. V.R.Manfredi,
 Dipartimento di Fisica, via Marzolo 8, 35131 Padova (Italy); \hfill\break
\noindent
e--mail: MANFREDI@PADOVA.INFN.IT or VAXFPD::MANFREDI}

\vskip 2 truecm

\centerline{M.Ferlini$^{(+)}$,

V.R.Manfredi$^{(+)(\ddagger)(\bigtriangleup)}$,  G.Maino$^{(\circ)}$ }

\vskip 1 truecm

\noindent
\item{$^{(+)}$} Dipartimento di Fisica G. Galilei

\item{$^{(\ddagger)}$} INFN, Sezione di Padova

\vskip 1 truecm

\item{$(\triangle)$} Interdisciplinary Laboratory, SISSA, Italy
\item{$(\circ)$} ENEA, Bologna, Italy and INFN, Sezione di Firenze

\vskip 2 truecm

\noindent  {\bf ABSTRACT}

\noindent
  To show the existence of precursor phenomena of the transition order$\
to$chaos in atomic
nuclei a simple analysis has been made, based on a recent criterion
proposed by Pavli\-chenkov.
The basic idea is that nonlinear effects in rotational bands should lead to
bifurcation, which qualitatively changes the spectrum for one of the
nuclear excitation branches.

Experimental data concerning the isotopes of Er, Hf,  Yb, Os, Dy, W
strongly
support the above
criterion.

\vskip 1 truecm


\vfill\eject

\noindent{\bf  1.  Introduction}

Many experimental and theoretical results strongly support the coexistence
in atomic nuclei of regular and chaotic motions [1, 2, 3, 4]. In this
respect it
is of
great interest to study the transition region between the two regimes, in
order to investigate the possible existence of some precursor phenomena of
the
full chaotic motion. As
stressed by the author of [5], nonlinear effects in rotational spectra
should
lead to bifurcation which qualitatively changes the spectrum for one of the
nuclear excitation branches [6, 7, 8]. Nonlinear effects, such as
centrifugal
and
Coriolis forces, can produce a qualitative change in the energy spectrum
and
may be considered a signature of the transition mentioned above. In
particular the Coriolis force arises from the coupling between
single--particle and rotational motions and this force manifests itself in
the
rotational spectra of odd nuclei, particularly where an odd nucleon
occupies
the anomalous parity subshell states with a high angular momentum j
($i_{13/2}$ for neutrons and $h_{11/2}$  for protons in rare--earth nuclei)
[5].
In reference [7] and [8], it was shown that bifurcation can be observed in
rotational bands based on these states for the Yb odd--N nuclei.

The aim of this work is to show that the same behaviour occurs for
different

nuclei, such as the isotopes of Er, Hf,  Yb, Os, Dy, W.[9].

The structure  of the article is as follows: in section 2 the particle--
rotor
model is  briefly discussed; in section 3 the numerical results are
shown.

\bigskip
\noindent {\bf 2. The Particle--Rotor Model}

This model has been discussed in great detail in references [5, 7, 8] and
in
this
section we report only some basic formulae. Following the notation of
reference [8] the Hamiltonian of this model can be  written:

$$ H= {R^2 \over 2J} +Q j^2_3. \eqno (2.1) $$

\noindent
The first term is the rotational energy of the core with moment of
inertia $J$ and the second term  describes the splitting of subshell levels
as
a consequence of the  quadrupole deformation:

$$ Q= {(N+3/2)\hbar \omega_0 \delta \over 2j(j+1) }(u^2_j - v^2_j), \eqno (
2.2)
$$

\noindent
where $\delta$ is the deformation parameter and $j$ the single particle
angular
momentum.
The total angular momentum, $I$, is thus given by the sum $\vec R + \vec j$
, and
$j_3$ denotes
the projection of $\vec j$ on the symmetry axis of deformed nucleus, see
fig. 1.
Moreover, eq.
(2.2) contains the amplitudes  $u_j$ and $v_j$,  depending on the subshell
occupation and $Q$ may take  both positive and negative values.

The energy levels of the Hamiltonian (2.1) can be grouped into rotational
multiplets with the same  $I$. If we define as {\it
signature} $\sigma = (-1)^{I-j}$,   {\it favoured band} $(\sigma=+1)$ and
{\it unfavoured band}  $(\sigma=-1)$ show a different behaviour with
respect to
the 180$^\circ$  rotation about axis 2 (see figure 1).

Let

$$ I_c = j (1-p+ \sqrt{p^2 -2p}), \quad\quad  p=QJ  \eqno (2.3) $$

\noindent
be a critical angular momentum value,
for $I<I_c$ the levels of the favoured and unfavoured bands come closer to
one another. For further discussion of this point, see [5], [7].

A rough estimation  of the amplitudes $u_j$ and $v_j$ can be obtained in
the framework of the Bardeen--Cooper--Schrieffer (BCS)  superconducting
model [10]. The wave function of a nuclear system with an even number of
particles is thus given by

$$ \vert \psi_e > = \prod_{j>0} (u_{j} + v_{j} a^+_j
a^+_{\overline j}) \vert 0>, \eqno (2.4) $$

\noindent
while

$$ \vert \psi_0 > = a^+_{j_1} \prod'_{j>0} (u_j + v_{j}
a^+_j a^+_{\overline j} ) \vert 0>, \eqno (2.5) $$

\noindent
holds for an odd--nucleon system. Operator $a^+_j$ creates a nucleon  in
the
j--th
Nilsson level, characterized by positive projection of the
angular momentum on the symmetry axis.   $a^+_{\overline j}$  is the
corresponding
operator for the time--reversed state at the same energy, $\varepsilon_j$,
but
with negative spin projection.  In the products (2.4) and (2.5) only
positive

values of $j$ are present and the prime denotes that the $j_1-$th Nilsson
level, occupied by the  unpaired nucleon, must be dropped from the
product.  The whole single--particle space is then generated by the set of
Nilsson levels $\{j, \overline j\}$. In eqs. (2.4) and (2.5) $u_j$
and $v_j$ are variational parameters to be determined on the requirement
that
the
energy has a minimum; their squared values represent the probability that a
pair
state,
$(j, \overline j)$ in the Nilsson potential is or is not
occupied. Of course, one has

$$ \vert u_j \vert^2 + \vert v_j \vert^2 =1. \eqno (3) $$

By suitable assumption [10] for the pairing interaction and  for the phase
factor of the
nuclear wave functions (2.4) and (2.5), otherwise undetermined, $u_j$ and
$v_j$ can be defined real and positive, and evaluated in the usual way [10]
by
introducing the Bogoliubov--Valatin (BV) transformations from particles to
quasiparticles.   The resulting BCS equations for non--interacting
quasiparticles must be solved iteratively, since they are nonlinear, to
obtain

the occupation probabilities for all the Nilsson levels, the Fermi energy,
$\lambda$, and the pairing gap, $\Delta$, for both neutrons and
protons.

The Fermi energy corresponds, mathematically, to a Lagrange multiplier
which fixes the expectation value of the particle number to the given
$N$ (number of neutrons or protons), respectively, since the BV

transformations, as is well known,  do not conserve it. However, the fact
that
the average
nucleon number in the ground--state has the correct value is sufficient for
our purposes, since we do not aim to describe in detail any one  particular
nuclear spectra,  but only to give a rough estimation of $I_c$. Thus  a
projection method
is unnecessary, which greatly simplifies
the computational procedure.

On the other hand, due to the breaking of the spherical rotational
symmetry,
the total
angular momentum is not a good quantum number and, to restore the
conservation
of the total
angular momentum, an analogous projection technique would have to be
adopted
[10].

With these drawbacks,  in the  present calculations of single--particle
levels,
we utilized
the Nilsson parametrization of ref. [11],
 and a pure pairing force in
the BCS  equations, whose strength has been determined by a semiempirical
analysis of mass differences for neighbouring nuclei [12].

  \vfill\eject

\noindent{\bf 3.  Analysis of the Experimental Data  }

In  this section we show the bifurcations in the rotational bands mentioned
in
section 2 for the isotopes  Dy, Er, Yb, Hf, W, Os.

For the sake of completeness, the Yb odd--N nuclei have also been shown, as
previously discussed in references  [5], [7].

The quantity

$$ \eta(I) = {E(I) - E(j) - A(I-j)^2 \over (2I+1)A} \eqno (3.1) $$

\noindent
is plotted as a function of $I$, where $E(I)$ is the energy of the level.
The parameter $A= \displaystyle{{1 \over 2J}}$ was determined  by linear
regression from
the experimental data of favoured bands.

In fig. 8 we also report  the experimental data for the Hf isotopes,
without the
use of
(3.1). Obviously the  bifurcation effect, discussed above, is less
pronounced,
but the
advantage of this analysis is that it is completely model independent.

In table I the main quantities used in the calculation of $I_c$ are shown.
As
can be seen, the
parameter $Q$ varies in an apparently random faschion. This effect is due
to the
factor
$u^2_j-v^2_j$, which changes very rapidly from one nucleus to another.

In spite of the crude model used, there is a rough agreement between the
calculated $I_c$
values (Th) and those extracted from the experimental data (Exp).

\vfill\eject

\noindent {\bf Conclusion}

Using  a simple approach, suggested by the author of [5, 7, 8], an analysis

was made  to show  the existence of bifurcations in the spectra of some
isotopes

(Er, Hf, Yb, W, Dy, Os).  As discussed above, this behaviour might be
connected
to
the transition order$\to$chaos present in atomic nuclei [1].

\bigskip
\centerline{***********}
\bigskip\noindent
One of us (V.R.M.) is greatly indebted to Prof. S.Fantoni, Director of the

Interdisciplinary Laboratory, SISSA (Trieste), for his kind hospitality in
his
lab.,
where the final version of this paper was written.

\noindent {\bf References}

\item{[1]} O.Bohigas and H.A.Weidenm\"uller: Ann. Rev. Nucl. Part. Science
{\bf
38}, 421 (1988).

\item{[2]} J.D.Garret, J.R.German, L.Courtney and J.M.Espino: in
\underbar{``Future Directions in} \hfill\break
\noindent\underbar{Nuclear Physics with $4\pi$ Gamma Detections Systems of
the
New Generation''}, ed. by J.Dudek and B.Haas (A.I.P. 259, N.Y.), p.345 (
1992).

\item{[3]} S. \AA berg: Prog. Part. Nucl. Phys. {\bf28}, 11 (1992).

\item{[4]} M.Matsuo, T.D\o ssing, E.Vigezzi and R.A.Broglia, Phys. Rev.
Lett.
{\bf 70}, 2694 (1993).
\item{} M.Matsuo, T.D\o ssing, B.Herskind, S.Frauendorf, E.Vigezzi, R.A.
Broglia:
Nucl. Phys. A {\bf 557}, 211c (1993).
\item{} B.R.Mottelson: Nucl. Phys. A {\bf 557}, 717c (1993).

\item{[5]} I.M.Pavlichenkov: Sov. Phys. JETP {\bf 69} (2), 227 (1989).

\item{[6]} I.M.Pavlichenkov and B.I.Zhilinskii: Ann. of Phys. (N.Y.) {\bf
184}, 1 (1988).

\item{[7]} I.M.Pavlichenkov: Phys. Rep. {\bf 226}, 173 (1993).

\item{[8]} I.M.Pavlichenkov: Europhys. Lett. {\bf 9} (1), 7 (1989).

\item{[9]} V.R.Manfredi and M.Ferlini: Bollettino della SIF, LXXVIII
Congresso
Nazionale, 189 (1992).

\item{[10]} P.Ring and P.Schuck, The Nuclear Many--Body Problem,
Springer--Verlag, New York (1980) chapt. 6.

\item{[11]} P.A.Seeger and W.M.Howard, Nucl. Phys.  {\bf A 238} (1975) 491.

\item{[12]} P.E.Nemirovsky, Kurchatov Report IAE--2530 (1975).

\vfill\eject

\noindent {\bf Figure Captions}

\item{} Fig. 1: The precession of the vectors $I$, $j$ and $R$ in the
vicinity
of
the critical point $I_c$; adapted from reference [5].

\item{} Fig. 2: The onset of  bifurcation in the rotational spectra of the
odd

isotopes of Er; \hskip 2 truecm {\it favoured band}, \hskip 2 truecm {\it
unfavoured band}.
\item{} Fig. 3: The same  as  Fig. 2 for the odd isotopes of Hf.

\item{} Fig. 4: The same  as  Fig. 2 for the odd isotopes of Yb.

\item{} Fig. 5: The same   as Fig. 2 for the odd isotopes of W.

\item{} Fig. 6: The same  as  Fig. 2 for the odd isotopes of Dy.

\item{} Fig. 7: The same  as Fig. 2 for the odd isotopes of  Os.

\item{} Fig. 8:  Experimental data for the Hf isotopes, without the use of
(3.1); the energy
is expressed in keV.

\vfill\eject

\noindent {\bf Table Captions}

\noindent
Tab. 1: The most important quantities used in the calculation of $Q$ and
$I_c$.
$j$: single
particle angular momentum; $\delta$: deformation parameter; the other
quantities
are
defined in the text.

\bye